# Reading Stockholm Riots 2013 in social media by text-mining

Andrzej Jarynowski[1,2,3], Amir Rostami[1]

1) Department of Sociology, Stockholm University, Stockholm, Sweden
2) Department of Theory of Complex System, Smoluchowski Institute, Jagiellonian University, Cracow, Poland
3) Laboratory of Technics of Virtual Reality, CIOP – National Research Instutute, Warsaw, Poland
andrzej.jarynowski@sociology.su.se

**Abstract**

The riots in Stockholm in May 2013 were an event that reverberated in the world media for its dimension of violence that had spread through the Swedish capital. In this study we have investigated the role of social media in creating media phenomena via text mining and natural language processing. We have focused on two channels of communication for our analysis: Twitter and Poloniainfo.se (Forum of Polish community in Sweden). Our preliminary results show some hot topics driving discussion related mostly to Swedish Police and Swedish Politics by counting word usage. Typical features for media intervention are presented. We have built networks of most popular phrases, clustered by categories (geography, media institution, etc.). Sentiment analysis shows negative connotation with Police. The aim of this preliminary exploratory quantitative study was to generate questions and hypotheses, which we could carefully follow by deeper more qualitative methods.

## 1. Introduction

The language and its products are the most important source of information of people's states of mind in communication process (Babtist 2010). The semantic and structural analysis of texts could shed more light on human perception, interpretation and creation of social phenomena like riots. Moreover, thinking in terms of networks and hierarchy (associations between words and terms) has gained ground in many disciplines (Mouge 2003), including psychology, criminology, anthropology, political science, while it has risen from physics and sociology. Our aim is to provide qualitative analysis of how people discuss Stockholm riots and what are their stimuli to such activity. To do so, we investigate Forum posts (Poloniainfo.se) and Twitter during and after riots. We choose those datasets, because, they are freely available and can be legally crawled from Internet and both can be treated as big data. Unfortunately, both Medias represent bias, due to very special category of people using them, and conclusions based on those communities cannot be directly generalize for the whole population. However, some insights for future studies could be obtained, especially from Twitter (Polonia.info is even less representative and findings from there could only help us to understand the complexity of riot phenomena). The scientific community has already experienced the power of social network media since the riots in Tottenham, north London, in August 2011 (theguardian&LSE 2011). Since then Twitter and any other social media were carefully investigated for almost every social movement such as STOP-ACTA (Jarynowski 2013) or Smoleńsk crash (Sobkowicz 2013) in Poland with computational tools constructed for this problem. We try some simple tools of NLP (natural language processing) and text-mining (Yuskiv 2006) to obtain some kind of hierarchical or network structure of concepts mentioned by Internet-users. The initial disturbances in Husby, a suburb in the northern part of Stockholm, were triggered by the police shooting of an old Maghreb-origin man. Most of the discussion took place during riots since 15.05 (incident with police) via period of actual riots 20-25.05 and shortly after that. Both Forum and Twitter data were collected from 15.05 till middle of July (15.07) so both data series are exactly 2 months long.

With the relatively high standard of living in Sweden (even suburban area of Stockholm like Husby is a long way away from poorness), the economic grievances were probably not the primary motivating factor in these protests (Barker 2013). The riots in Stockholm have been carried out by a combination of angry local youths, radical left-wing activists and hardened criminals (Rostami 2012). Swedish Police have already intervened against many similar size riots like in Rosengård (Malmö) in 2009 or in other neighborhoods often gathered people with socio-economic problems (Nilson 2011). Criminologists note that the segregation issues, which have been driving riots in 20$^{th}$ century, turn in 21$^{th}$ into: alienation, rootlessness, unemployment, distrust and resentment against society (Hallsworth 2011). Our goal is longitudinal text-mining analyses of public opinion (within social media) on the riot in order to explain its phenomena theoretically described by criminologists, put some light on driving factors, which warm up the discourse and set up some hypothesis about emergence of media phenomena.

## 2. Twitter

### 2.1. Data description and objectives

We analyze ~14k Tweets in different languages (mostly Swedish and English) tagged with hash Husby. That's implying international perspective of people, who express their thoughts via Twitter. Because of Multilanguage perspective of such Tweets, we decide to analyze not whole content of those tweets, but only co-occurrence with other hash tags. We do not differentiate who was twitting: simple users, mainstream media, non-mainstream media, bloggers, activists, or even and the police. Only ~ 8k Tweets were taken in our analysis (only those with more than one hash tag). We choose the 20 most frequent tags, there Svpol is the most frequent with Sthlmriots and Migpol far behind and rest (Table 1). We decided to analyze hash tags, just as they are, but there could be possibility to categorize some hash tags in just one category (e.g. by combining Sthlmriots with Stkhlmriot).

| No | Hash tag | English meaning | Counts |
|---|---|---|---|
| 1 | Svpol | Swedish Politics | 3897 |
| 2 | Sthlmriots | Sthlmriots | 1319 |
| 3 | Migpol | Migration Politics | 436 |
| 4 | Sthlmriot | Sthlmriot | 236 |
| 5 | Stockholm | Stockholm | 200 |
| 6 | Aftonbladet | Press company | 142 |
| 7 | Nymo | Press company | 124 |
| 8 | Rinkeby | District of Stockholm | 109 |
| 9 | Polisen | Police | 108 |
| 10 | Sweden | Sweden | 100 |
| 11 | Upplopp | Riots | 92 |
| 12 | Kista | District of Stockholm | 89 |
| 13 | Svtdebatt | Debate in Swedish TV | 82 |
| 14 | Vpol | Left Politics | 80 |
| 15 | Debatt | Debate | 76 |
| 16 | 08pol | Police PR Department | 75 |
| 17 | Expressentv | TV Program | 72 |
| 18 | Megafonen | Political Activists | 71 |
| 19 | Kravaller | Riots | 70 |
| 20 | Tensta | District of Stockholm | 69 |

Table 1: Most frequent tags

### 2.2. Longitude analysis

We found, that for example Svpol (Fig. 1) and Migpol (Fig. 2) are tags, which were in constant use for the whole period. The shape of cumulative hash tags counts curve for them is almost linear. That means: the occurrence probabilities are equal in time. Moreover, every second twit since beginning of riots till middle of July is associated with Svpol hash tag. The rest of the hash tags died out after riots finished.

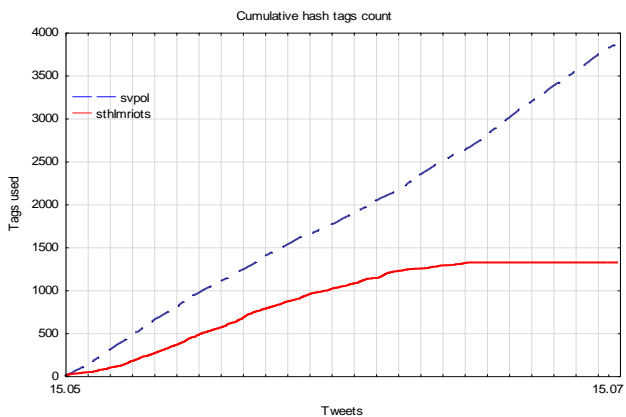

Fig. 1: Longitudinal analysis of two most frequent tags (Svpol, Stkhlmriots)

Moreover, for tags like Debatt or Svtdebatt, we observed, that people used them only around the event (Fig. 2), which is a very common phenomena in Twitter world. Some media names hash tags have a stepwise shape like Nymo (Fig. 2), which also is characteristic for media. Those institutions provide some news, which are likely to be retwited. That explains the big number of media hash tag use in short time surrounded by quiet regions (Chmiel 2011). The frequency of usage of given media hash tags could be also an indicator of how influential that medium is.

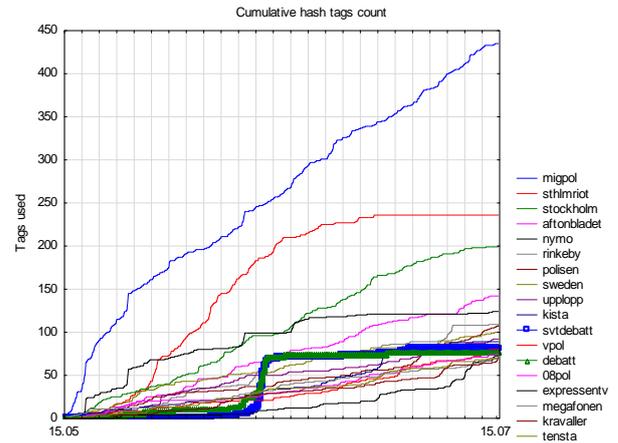

Fig. 2: Longitudinal analysis of rest of most frequent tags

### 2.3. Association analysis

We also try to find assotiation between tags. We define the link, when in the same tweet both tags coexist. Hierarchical analysis shows leading role of dyad Svpol-Sthhlmriots (Fig. 3, Table 2). Moreover, 2-gram elements (co-occurrence of 2 terms in one twit) of main dual dyad sthlmriots_svpol and svpol_migpol are a few times more frequent than other elements (Table 2). However, we cannot call triangle Migpol, Svpol, Sthlmriots as triad, because the link between Migpol and Sthlmriots was observed only 37 times so it is an order of magnitude weaker than main double dyad.

| No | Hash tag | Counts |
|---|---|---|
| 1 | sthlmriots_svpol | 533 |
| 2 | svpol_migpol | 353 |
| 3 | svpol_sthlmriot | 107 |
| 4 | migpol_nymo | 50 |
| 5 | vpol_svpol | 47 |

Table 2: The most frequent 2-grams. Evidence of importnace of dual dyad sthlmriots_svpol and svpol_migpol.

Let's define construction of the network (Fig. 3). We decided to establish lower threshold on the level of 2 tweets, needed to create a link (everything below seems to be a noise, because link- association should be repeated at least once to avoid random effects). Thickness of the link corresponds to its weight (count of given 2-gram). On the other hand, dual dyad sthlmriots_svpol and svpol_migpol thicknesses were reduced not to cover the whole figure.

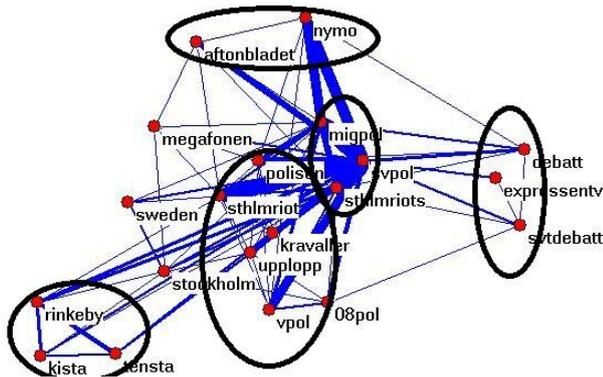

Fig. 3: Network of connections (links thinness in dual dyad sthlmriots_svpol and svpol_migpol were decreased to see other links also)

Network analysis (Buda 2013) show leading role of dyad Svpol-Sthhlmriots and quasi-triangle Migpol, Svpol, Sthlmriots. On the other hand we can find clusters of geographical districts, tags related to debate, Swedish words describing riots, and media institution.

### 2.4. Conclusions of Twitter analysis

The provided analysis shows many features known already from other studies (stepwise functions for cumulative Tweets count for media organization or clusterization of hashes within similar semantic field) and general observation, but here they are presented in a more systematized way. The most important topic is politics. Svpol as other hashes with the same meaning, are definitively the most frequent hash tags. Moreover only Svpol and Migpol seem to be used after riots with the same frequency as before. It would be interesting to see how hashes about politics co-occurrence with others change with time. Another question could be asked with sentiment analysis: how emotionally oriented are hashes about politics.

## 3. Poloniainfo.se

### 3.1. Data description and objectives

Internet Forums like poloniainfo.se are not broadcasting media as Twitter, but relations between users are usually stronger and more personal. Quantitative research could be deeper due to complex relations between users (Zbieg 2012), but on the other hand amount the of data is not as impressive as Twitter. We look at frequencies of world used by Forum user in Topic about Riots in 525 Posts. We choose only Polish words in this analysis. Firstly we found extremely huge amount of personal and possessive pronouns of third person in plural form (Table 3). Everything seems to be describing about "Them" more often called "others" in ethnological literature (Bauman 1996). "They" are native Swedes represented mostly by government and police and another "They": riot participants. That indicates observative and little biased way of looking on the riots presented by Forum users (Gustafsson *in press*). The Polish community did not take part in riots, and on the other hand have no significant influence on politics of Sweden. That makes this medium neutral, while views: both for and against the riots were presented there. However, Polish people identify themselves culturally with Swedish establishment and describe problems of Husby citizens unlike their own perspective.

| interesting pronoun | Polish | freq | compared pronoun | Polish | freq |
|---|---|---|---|---|---|
| **them** | **im** | **79** | us | nam | 4 |
| **the** | **tym** | **102** | us | nam | - |
| **them/ their/ theirs** | **ich** | **72** | ours/ our | nasze/ nasz/ nasza | 2 |
| **they** | **oni** | **53** | we | my | 1 |
| **these** | **ci** | **48** | we | my | - |
| **them** | **nich** | **38** | us | nas | 13 |

Table 3: Orientation of conversation on „them"

### 3.2. Methodology and data mining

We tried to categories words used in discussion in few categories. To do so, we choose only those words, which have only one clear meaning somehow related to the topic. We found 386 different words, which appear at least once in our sample and seem to have some important meaning. From them around 300 were attached to different categories 1-10 with subcategories described by some keywords (Table 4). Every category allocate sum of number of unique words, which belong to family of given keywords related to given category or subcategories.

| |
|---|
| 1.1 Employment (work/employees, hardworking, rich, money, taxes) |
| 1.2 Unemployment (unemployment, social help, poor) |
| 2 Family (family) |
| 3 Religion (Islam, religion) |
| 4 Education (education, schools, learn, language) |
| 5 Living (apartments/residents, district) |
| 6.0 Politics-general (government, debate, party, politicians, democracy) |
| 6.1 Politics-multikulti (invite, acclimatization, multikulti politics, hope, tolerance, asylum, get, arrives) |
| 6.2 Politics-segregation (racist names, eugenics, racism, segregation, deportation, hate) |
| 7.0 Identity-general (nation, Stockholm, society) |
| 7.1 They (immigrants, Arabian, other nations, origin) |
| 7.2 Swedes (Swedes, Swedish, Sweden, Europe, nobility) |
| 8.0 Police-general (police, military) |
| 8.1 Police-induce (killed, wounds, induce, bullets, weapon, shoot, Police in slang, knife, disarm) |
| 8.2 Police-law (law, cutthroat, action in the name of law) |
| 9.0 Riots-general (throw, riots, night, street, violence, stones, car) |
| 9.1 Riots-pro (rebellion, youth, protest, vulnerable) |
| 9.2 Vandalism (fires, vandalism, aggression) |
| 10 External fields (other riots, problems, wars, media) |

Table 4: Categories, subcategories and keywords describing them

## 3.3. Limitation of coding

Meaning of words used by people is very difficult to uniqueness classification. In our task we propose 10 main categories with 14 subcategories and attract presented words into given keywords related to descriptive category or subcategory. Classification is based on our subjective feeling. We tried to avoid words with many meanings. We had problems with words:
- Stockholm/Sweden (does not only relate with that city/country, but also geographical location);
- Swedish/Swedes (does not only describe citizenship of Sweden, but also the background of the riots);
- all words classified to categories pro riots or pro police (words connoted with law or rebellion have mostly positive meaning, but not always);
- all words classified to category external forces (e. g. none of media institution like radio, TV, press, or Internet companies names were included in investigation)

Also all of categories have very wide range of potential connotation and some of them could overlap (e. g. where should be the border between employment and unemployment), but we tried to help ourselves with keyword list (Table 4).

## 3.4. Results of categorization

From results, we can conclude, that the main conflict is going on around identity, police operation and riots itself, work and living issues. Politics, education and family related issues plays secondary role, but still such topics were discussed by Forum users. The main subject of discussion seems to focus around identity (the biggest count of related words) which was already observed from intensively of pronounce use (Table 3). Moreover, motor of conflict could be defined as Swedes-They. While there are some coding problems with identity (are literary Swedes associated with literary category 7.2?), the second frequent category: Employment is probably the biggest single issue related to Stockholm's riots mentioned by Forum users. The smallest subcategory (Police-pro) is one related to the positive side of police operation. One order of magnitude often Police was described by negative or neutral connotation.

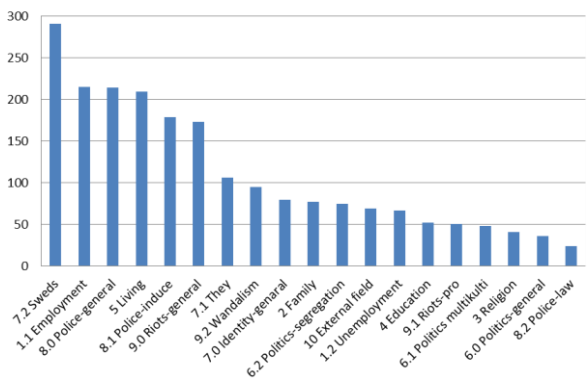

Fig. 4: Categories and Subcategories from most to least popular

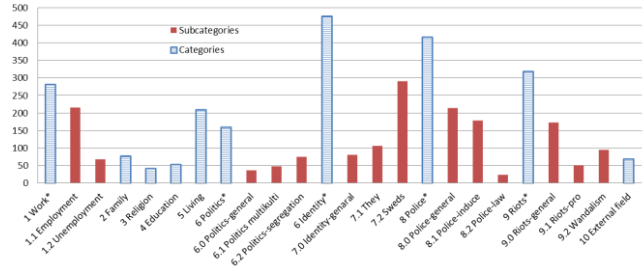

Fig. 5: Categories and categories containing subcategories. * means sum of all counts of subcategories for given category

## 3.5. Sentiment analysis of 2-grams with Police

We extract all 2-grams where the word "police" or "police officer" appears. We run sentiment analysis on each 2-gram. The sentence sentiment strength[1] could vary from -4 (very negative) to +4 (very positive). Most of them have been neutral and sensitive strength is 0. We analyze all those 2-grams, which were found more than once. To illustrate, we show few most frequent 2-grams (Table 5). Moreover, Polish stop words (without meaning) were excluded also. To get effective power, we multiply frequency by sensitive strength. The overall score of power for all 2-grams containing word Police is slightly negative (-5).

| Polish 2-gram | English 2-gram | No. | sensi strenth | Pow. |
|---|---|---|---|---|
| szwedzka policja | Swedish police | 10 | 0 | 0 |
| policja używa | police uses | 3 | 0 | 0 |
| granatniki policjanci | police launchers | 2 | -1 | -2 |
| jaka policja | what the police | 2 | 0 | 0 |
| mogła policja | police could | 2 | 0 | 0 |
| mordować policja | police murders | 2 | -3 | -6 |
| … | … | … | … | … |
| | | | Sum Power | -5 |

Table 5: Most frequent 2-grams with Police, Policeman or Police officer with their sentiment score and frequency.

## 3.5. Conclusions and future works for Poloniainfo.se

This work has only exploratory function, but even those preliminary results; we can propose few hypotheses, which should be checked by deeper investigation.
The first one is related to Police operation. Please note, that we do not want to evaluate professionalism of Police, but only public opinion about their operation. Data-mining analysis proposed by us has many weak sides and it is usually very difficult to make clear conclusions from it. In the Police case, order of magnitude difference

---
[1] We use tool: Sesnistrenth (Thelwall 2012) for English translation of Polish sentences

between positive and both negative and neutral cases seem to be something more than methodological bias or artifact. Why Forum users even, they identify themself with Swedish establishment did not say almost anything positive about Police? Did PR department of Police work properly? One user wrote, that Swedish Police as a best paid organization in EU is one of the less effective at the same time. We propose to survey public evaluation about Police operation. Moreover sentiment and association analysis should also give some more insight, while preliminary results show little negative emotional orientation (Table 5). This sample is unfortunately too small for any conclusions and bigger datasets should be used to estimate actual sensitive power.

Employment also seems to be relatively important topic (Fig. 4, 5). Work issue, with connotation with taxes and salaries should be more carefully investigated if that is really a leading factor of discussions about riots. It beats some aspects of identity (and even identity also if other coders would not include geographical word – e. g. Stockholm into identity category) and religion, living condition or education form frequentional analysis. However, it could come from bias, that Poles describing riots are mostly guest workers and work, as a single theme, is the most popular within Polish community in Sweden.

## 4. General findings, limitation of both studies and speculations

The role of Police and politics should be investigated, because in Twitter politics (Table 2) and in Forum Police (Fig. 5) are the main motors of opinion spread between people. Both studies are only preliminary and just explore the field to set up questions to be ask by pragmatic research. Both datasets are not representative for the whole society and opinion shared in both mediums are very special to people who use them. Moreover in each study different methodology was used due to difference of data structure itself, and even in content. Some aspects revealed in one study were omitted in the second one. For example media institutions, which play important role in hash tag study, were not categorized anywhere in Forum case. However, similar issues come out from both mediums. As the result from this preliminary analysis hypotheses about emergence of media phenomena appear, while similar size riots reported by police went unnoticed (Nilsson 2012). For example: how influential actors used social media like bloggers, journalist, political activist made this topic so popular and it has been discussing till now (Fig. 1). The next part of this project will focus on network analysis of role of such actors.

While sociologists try to understand mechanisms of riots arising, underlying psychological and sociological patterns, and the presentation of riots in the mass media, natural language processing and text mining are great supplementary tools for that. Criminologists, on the other hand, focus on recent causation theories and suggest various ways of controlling riots (Sarnecki 2001). They are aware of similar incidents and forecast intensification of riots in the future due to stratification of society with overrepresentation of "urban underclass". That indicates the need for adjusting computational methods to problems, and this paper is a preliminary approach for that.

## Acknowledgments

We would like to thank Hernan Mondani, Andrzej Grabowski, Fredrik Liljeros, Anita Zbieg, Lauren Dean, and Clara Lindblom, for discussions. AJ thanks to Svenska Institutet for invitation to Sweden.

## References

Barker, V. (2013). Policing Membership in Husby: Four Factors to the Stockholm Riots. *Border Criminologies Blog Post*: 3 June.

Babtist, J.M. (2010.) Quantitative Analysis of Culture Using Millions of Digitized Books. *Science* 331 (6014).

Bauman, Z. (1996). *From pilgrim to tourist–or a short history of identity. Questions of cultural identity*. pp 19-38. London.

Buda, A. and Jarynowski, A. (2013). Network Structure of Phonographic Market with Characteristic Similarities between Artists, In: *Acta Physica Polonica A*, vol. 123 (2).

Chmiel, A., Sienkiewicz, J., Thelwall, M., Paltoglou, G., Buckley, K., Kappas, A., & Hołyst, J. A. (2011). Collective emotions online and their influence on community life. *PloS one*, *6*(7), e22207.

Gustafsson, S.M., Sikström, S. & Lindholm, T. (in press). Selection Bias in Choice of Words: Evaluations of "I" and "We" Differ between Contexts, but "They" are Always Worse. *Journal of Social Psychology and Language.*

Hallsworth, S. and Brotherton, D. (2012). Urban Disorder and Gangs: A Critique and a Warning. London: Runnymede Trust.

Jarynowski, A., Zbieg, A. and Jankowski, J. (2013) Viral spread with or without emotions in online community. *arXiv preprint arXiv:1302.3086*.

Mouge, P. & Contractor, N. (2003). *Theories of Communication Networks.* Cambridge: Oxford University Press.

Nilsson, T., and Westerberg, A. I. (2012). *Våldsamma upplopp i Sverige–från avvikelse till normalitet.* MSB:Stockholm.

Theguardian & LSE. (2011). *Reading the Riots*. London.

Thelwall, M. (2012) Sentiment strength detection for the social Web. *Journal of the American Society for Information Science and Technology,* vol. 63 (1).

Yuskiv, B. (2006). *Content analysis. History development and world practice*. Rowne.

Rostami, A. (2013). *Tusen Fiender*, Linnéuniversitetet:Linkopping.

Sarnecki, J. (2001). *Delinquent networks*. Cambridge: Cambridge University Press.

Sobkowicz, P. and Sobkowicz, A. (2012). Two-year study of emotion and communication patterns in a highly polarized political discussion forum. *Social Science Computer Review*, *30*(4), 448-469.

Zbieg, A., Żak, B., Jankowski, J., Michalski, R. and Ciuberek, S. (2012). Studying Diffusion of Viral Content at Dyadic Level. In: ASONAM 2012, Istambul.